\documentclass[11pt]{article}
\usepackage{amssymb}
\usepackage[dvips]{lscape,graphicx}

\newcommand{\bc}{\begin{center}}
\newcommand{\ec}{\end{center}}
\newcommand{\bd}{\begin{displaymath}}
\newcommand{\ed}{\end{displaymath}}
\newcommand{\be}{\begin{equation}}
\newcommand{\ee}{\end{equation}}
\newcommand{\ba}{\begin{array}}
\newcommand{\ea}{\end{array}}
\newcommand{\bt}{\begin{tabular}}
\newcommand{\et}{\end{tabular}}

\newcommand{\ds}{\displaystyle}

\begin{document}

\hyphenation{ }

\title{No--scale supergravity and the Multiple Point Principle}

\author{ C.Froggatt${}^{1}$, L.Laperashvili${}^{2}$, R.Nevzorov${}^{3,2}$, H.B.Nielsen${}^{4}$\\[5mm] 
\itshape{${}^{1}$ Department of Physics and Astronomy, Glasgow University, Scotland}\\[0mm] 
\itshape{${}^{2}$ Theory Department, ITEP, Moscow, Russia}\\[0mm] 
\itshape{${}^{3}$ School of Physics and Astronomy, University of Southampton, UK}\\[0mm]
\itshape{${}^{4}$ The Niels Bohr Institute, Copenhagen, Denmark}}

\date{}

\maketitle

\begin{abstract}
\noindent We review symmetries protecting a zero value for the
cosmological constant in no--scale supergravity and reveal the
connection between the Multiple Point Principle, no--scale and
superstring inspired models.

\end{abstract}

\section{Introduction}

Nowadays the existence of a tiny energy density spread all over
the Universe (the cosmological constant), which is responsible for
its accelerated expansion, provides the most challenging problem
for modern particle physics. A fit to the recent data shows that
$\Lambda \sim 10^{-123}M_{Pl}^4 \sim 10^{-55} M_Z^4$ \cite{0}. At
the same time the presence of a gluon condensate in the vacuum is
expected to contribute an energy density of order
$\Lambda_{QCD}^4\simeq 10^{-74}M_{Pl}^4$. On the other hand if we
believe in the Standard Model (SM) then a much larger contribution
$\sim v^4\simeq 10^{-62}M_{Pl}^4$ must come from the electroweak
symmetry breaking. The contribution of zero--modes is expected to
push the vacuum energy density even higher up to $\sim M_{Pl}^4$.
Thus, in order to reproduce the observed value of the cosmological
constant, an enormous cancellation between the various
contributions is required. Therefore the smallness of the
cosmological constant should be considered as a fine-tuning
problem. For its solution new theoretical ideas must be employed.

Unfortunately the cosmological constant problem can not be
resolved in any available generalization of the SM. An exact
global supersymmetry (SUSY) ensures zero value for the vacuum
energy density. But in the exact SUSY limit bosons and fermions
from one chiral multiplet get the same mass. Soft supersymmetry
breaking, which guarantees the absence of superpartners of
observable fermions in the $100\,\mbox{GeV}$ range, does not
protect the cosmological constant from an electroweak scale mass
and the fine-tuning problem is re-introduced.

It was argued many years ago that soft breaking of global
supersymmetry at low energies could be consistent with a zero
value for the cosmological constant in the framework of $N=1$
supergravity (SUGRA) models \cite{01}. Moreover there is a class
of models (so called no--scale supergravity) where the vacuum
energy density vanishes automatically \cite{34}. It happens
because no--scale models possess an enlarged global symmetry. Even
after breaking, this symmetry still protects zero vacuum energy
density at the tree level. All vacua in the no--scale models are
degenerate, which provides a link between no--scale supergravity
and the Multiple Point Principle (MPP) \cite{02}. MPP postulates
that in Nature as many phases as possible, which are allowed by
the underlying theory, should coexist. On the phase diagram of the
theory it corresponds to the special point -- the multiple point
-- where many phases meet. According to the MPP, the vacuum energy
densities of these different phases are degenerate at the multiple
point.

This article is organized as follows: in section 2 we describe the
structure of $(N=1)$ SUGRA models; in section 3 we study
symmetries protecting the zero value of the cosmological constant
in the no-scale models ignoring the superpotential; the no--scale
models with a non--trivial superpotential are considered in
section 4. The connection between the MPP, no--scale and
superstring inspired models is discussed in section 5.

\section{$N=1$ supergravity}

The full $N=1$ SUGRA Lagrangian \cite{34},\cite{21}
is specified in terms of an analytic gauge kinetic function
$f_a(\phi_{M})$ and a real gauge-invariant K$\ddot{a}$hler function
$G(\phi_{M},\bar{\phi}_{M})$, which depend on the chiral superfields
$\phi_M$. The function $f_{a}(\phi_M)$ determines the kinetic
terms for the fields in the vector supermultiplets and the gauge
coupling constants $Re f_a(\phi_M)=1/g_a^2$, where the index $a$
designates different gauge groups. The K$\ddot{a}$hler function is a
combination of two functions
\be
G(\phi_{M},\bar{\phi}_{M})=K(\phi_{M},\bar{\phi}_{M})+
\ln|W(\phi_M)|^2\, ,
\label{22}
\ee
where $K(\phi_{M},\bar{\phi}_{M})$ is the K$\ddot{a}$hler potential whose
second derivatives define the kinetic terms for the fields in the
chiral supermultiplets. $W(\phi_M)$ is the complete analytic
superpotential of the considered SUSY model. In this article
standard supergravity mass units are used:
$\ds\frac{M_{Pl}}{\sqrt{8\pi}}=1$.

The SUGRA scalar potential can be presented as a sum of $F$-- and D--terms
$V=V_{F}+V_{D}$, where the F--part is given by \cite{34},\cite{21}
\be
\ba{c}
V_{F}(\phi_M,\bar{\phi}_M)=e^{G}\left(\ds\sum_{M,\,\bar{N}} G_{M}G^{M\bar{N}}
G_{\bar{N}}-3\right)\, ,\\[3mm]
G_M \equiv\partial_{M} G\equiv\partial G/\partial \phi_M,
\qquad G_{\bar{M}}\equiv
\partial_{\bar{M}}G\equiv\partial G/ \partial \phi^{*}_M\, , \\[3mm]
G_{\bar{N}M}\equiv
\partial_{\bar{N}}\partial_{M}G=\partial_{\bar{N}}\partial_{M}K\equiv
K_{\bar{N}M}\, . \ea \label{6} \ee The matrix $G^{M\bar{N}}$ is
the inverse of the K$\ddot{a}$hler metric $K_{\bar{N}M}$. In order to
break supersymmetry in $(N=1)$ SUGRA models, a hidden sector is
introduced. It contains superfields $(h_m)$, which are singlets
under the SM $SU(3)\times SU(2)\times U(1)$ gauge group. If, at
the minimum of the scalar potential (\ref{6}), hidden sector
fields acquire vacuum expectation values so that at least one of
their auxiliary fields 
\be
F^{M}=e^{G/2}\sum_{\bar{P}}G^{M\bar{P}}G_{\bar{P}} \label{61} 
\ee
is non-vanishing, then local SUSY is spontaneously broken. At the
same time a massless fermion with spin $1/2$ -- the goldstino --
is swallowed by the gravitino which becomes massive
$m_{3/2}=<e^{G/2}>$. This phenomenon is called the super-Higgs
effect.

It is assumed that the superfields of the hidden sector interact
with the observable ones only by means of gravity. Therefore they
are decoupled from the low energy theory; the only signal they
produce is a set of terms that break the global supersymmetry of
the low-energy effective Lagrangian of the observable sector in a
soft way. The size of all soft SUSY breaking terms is
characterized by the gravitino mass scale $m_{3/2}$.

In principle the cosmological constant in SUGRA models tends to be
huge and negative. To show this, let us suppose that, the
K$\ddot{a}$hler function has a stationary point, where all
derivatives $G_M=0$. Then it is easy to check that this point is
also an extremum of the SUGRA scalar potential. In the vicinity of
this point local supersymmetry remains intact while the energy
density is $-3<e^{G}>$, which implies the vacuum energy density
must be less than or equal to this value. In general enormous
fine-tuning is required to keep the cosmological constant around
its observed value in supergravity theories.
\vspace{10mm}

\section{$SU(1,1)$ and $SU(n,1)$ symmetries in the no--scale models}

We know that the smallness of the parameters in a physical theory
can usually be related to an almost exact symmetry. Since the
cosmological constant is extremely tiny, one naturally looks for a
symmetry reason to guarantee its smallness in supergravity. In the
simple case when there is only one singlet chiral multiplet
$\hat{z}$, the scalar potential can be written as 
\be
V(z,\bar{z})=9e^{4G/3}G_{z\bar{z}}
\left(\partial_z\partial_{\bar{z}}e^{-G/3}\right)\,. 
\label{31}
\ee 
In order that the vacuum energy density of $V(z,\bar{z})$
should vanish, we must either choose some parameters inside $G$ to
be fine--tuned or, alternatively, demand that the K$\ddot{a}$hler
function $G$ satisfies the differential equation
$\partial_z\partial_{\bar{z}}e^{-G/3}=0$, whose solution is
\cite{31}: 
\be 
G=-3\ln\left(f(z)+f^{*}(\bar{z})\right)\,.
\label{32} 
\ee 
For the K$\ddot{a}$hler function given by
Eq.~(\ref{32}), fine--tuning is no longer needed for the vanishing
of the vacuum energy, since the scalar potential is flat and
vanishes at any point $z$. The kinetic term for the field $z$ is
then given by \be L_{kin}=\ds\frac{3\partial_z
f(z)\partial_{\bar{z}}f^{*}(\bar{z})}{(f(z)
+f^{*}(\bar{z}))^2}\left|\partial_{\mu}z\right|^2=
\ds\frac{3\left|\partial_{\mu}f(z)\right|^2}{(f(z)+f^{*}(\bar{z}))^2}\,.
\label{33} 
\ee 
As follows from Eq.~(\ref{33}), $L_{kin}$ can be
rewritten so that only the field $T=f(z)$ appears in the kinetic
term. Actually this holds for the whole Lagrangian. The considered
theory depends only on the field $T$ and all theories obtained by
the replacement $T=f(z)$ are equivalent.

One expects that such a theory with a completely flat potential
possesses an enlarged symmetry. For the case
$\mbox{$T=(z+1)/(z-1)$}$ the scalar kinetic term becomes
$$
L_{kin}=\ds\frac{3|\partial_{\mu}z|^2}{(|z|^2-1)^2}
$$
which is evidently invariant under the following set of
transformations: \be z\to\ds\frac{az+b}{b^{*}z+a^{*}}\,.
\label{34} \ee The set of transformations (\ref{34}) forms the
group $SU(1,1)$, which is non--compact and characterized by the
parameters $a$ and $b$ which obey $|a|^2-|b|^2=1$. Hence $SU(1,1)$
is a three--dimensional group. Transformations of $SU(1,1)$ are
defined by $2\times 2$ matrices
$$
U=\left(
\ba{cc}
a & b\\
b^{*}& a^{*}
\ea
\right)\,,
$$
which can also be written in the form \cite{32} 
\be
U=\exp\biggl\{i\frac{\omega_0}{2}\sigma_3+i\frac{\omega^{*}}{2}\sigma_{-}
-i\frac{\omega}{2}\sigma_{+}\biggr\}\,, \qquad
\sigma_{\pm}=(\sigma_1\pm i\sigma_2)/2 
\label{35} 
\ee 
Here $\omega_0$ is a real parameter and $\sigma_{1,2,3}$ are the
conventional Pauli matrices. The matrices $U$ acting on the space
$ \ds \left( \ba{c} x\\ y \ea \right) $ leave the element $|x|^2-|y|^2$ invariant, in
contrast with the $SU(2)$ group where we have invariance of the
element $|x|^2+|y|^2$\,. The $SU(1,1)$ transformations of the
field variable $T$ are
$$
T\to\frac{(\alpha T+i\beta)}{(i\gamma T +\delta)}\,
\qquad \alpha\delta+\beta\gamma=1\,,
$$
where $\alpha$, $\beta$, $\gamma$ and $\delta$ are real parameters.

The group $SU(1,1)$ contains the following subgroups \cite{33}:
\be
\ba{rll}
i)& \mbox{Imaginary translations:}&\qquad  T\to T+i\beta;\\[3mm]
ii)& \mbox{Dilatations:} & \qquad T\to\alpha^2 T ;\\[1mm]
iii)&\mbox{Conformal transformations:}&\qquad
T\to\ds\frac{\cos\theta T+i\sin\theta}{i\sin\theta T+\cos\theta}\,. 
\ea 
\label{36} 
\ee 
The K$\ddot{a}$hler function
(\ref{32}) is invariant under the first set of transformations,
but not under dilatations and conformal transformations. The
gravitino mass term in the SUGRA Lagrangian, which appears when
SUSY is broken, results in the breaking of $SU(1,1)\to U_a(1)$,
where $U_a(1)$ is a subgroup of imaginary translations. One can
wonder whether $SU(1,1)$ invariance implies a flat potential. The
invariance of the scalar potential with respect to imaginary
translations implies that $V(z,\bar{z})$ is a function of the sum
$z+\bar{z}$. At the same time the invariance under dilatation
forces $V(z,\bar{z})$ to depend only on the ratio $z/\bar{z}$.
These two conditions are incompatible unless $V(z,\bar{z})$ is a
constant. Moreover the $SU(1,1)$ invariance requires this constant
to be zero \cite{33}. In order to get a flat non--zero potential
in SUGRA models, one should break $SU(1,1)$. The $SU(1,1)$
structure of the $N=1$ SUGRA Lagrangian can have its roots in
supergravity theories with extended supersymmetry ($N=4$ or $N=8$)
\cite{34}.

Let us consider a SUGRA model in which there are $n$ chiral
multiplets $z$ and $\varphi_i$, $i=1,2,...n-1$, where $z$ is a
singlet field while $\varphi_i$ are non--singlets under the gauge
group. If the K$\ddot{a}$hler function has the form 
\be
G=-3\ln\left(f(z)+f^{*}(\bar{z})+g(\varphi_i,\bar{\varphi}_i)\right)\,,
\label{37} 
\ee 
then the F--part of the scalar potential vanishes
and only D--terms give a non--zero contribution, so that 
\be
V=\ds\frac{1}{2}\sum_{a}(D^{a})^2\,,\qquad
D^{a}=g_{a}\sum_{i,\,j}\left(G_i
T^a_{ij}\varphi_j\right)\,, 
\label{38} 
\ee 
where $g^a$ is the gauge coupling constant associated with the 
generator $T^a$ of the gauge transformations. Owing to the particular
form of the K$\ddot{a}$hler function (\ref{37}), the scalar potential
(\ref{38}) is positive definite. Its minimum is attained at the
points for which $<D^{a}>=0$ and the vacuum energy density
vanishes \cite{35}.

In the case when $g(\varphi_i,\bar{\varphi}_i)=-\sum_i\varphi_i
\bar{\varphi}_i$, the kinetic terms of the scalar fields are
invariant under the isometric transformations of the non--compact
$SU(n,1)$ group \cite{35}. The manifestation of the extended
global symmetry of $L_{kin}$ can be clearly seen, if one uses new
field variables $y_i$, i=0,1,...n-1, related to $f(z)$ and
$\varphi_i$ by
$$
f(z)=\frac{1-y_0}{2(1+y_0)}\,,\qquad \varphi_i=\frac{y_i}{1+y_0}\,.
$$
Then the K$\ddot{a}$hler function takes the form
$$
G=-3\ln\left(1-\sum_{i=0}^{n-1}y_i\bar{y}_i\right)
+3\ln|1+y_0|^2\,,
$$
from which it follows that the kinetic terms of the scalar fields
are 
\be
L_{kin}=\ds\sum_{j}\frac{3\partial_{\mu}y_j\partial_{\mu}\bar{y}_j}
{\left(1-\sum_{i}y_i\bar{y}_i \right)^2}\,. 
\label{39} 
\ee 
In particular the kinetic terms (\ref{39}) remain intact if 
\be
y_i\to \ds\frac{a_i y_i+b_i}{b_i^{*}y_i+a_i^{*}}\,;\qquad
y_j\to\ds\frac{y_j}{b^{*}_iy_i+a^{*}_i}\,\quad 
\mbox{for}\quad i\ne j\,, 
\label{310} 
\ee 
where $|a_i|^2-|b_i|^2=1$. The $SU(n,1)$ symmetry implies a zero 
contribution of the $F$--terms to the potential, which protects the 
vacuum energy density.

The $SU(n,1)$ symmetry can be derived from an extended ($N\ge 5$)
supergravity theory \cite{36}. This symmetry is broken by the
gauge interactions (D--terms) in $N=1$ supergravity models,
leaving only an $SU(1,1)$ symmetry. In terms of the symmetry
transformations (\ref{310}), the kinetic terms and scalar
potential are still invariant with respect to the replacement 
\be
y_0\to \ds\frac{a_0 y_0+b_0}{b_0^{*}y_0+a_0^{*}}\,;\qquad
y_i\to\ds\frac{y_i}{b^{*}_0y_0+a^{*}_0}\,\quad \mbox{for}\quad
i\ne 0\,. 
\label{311} 
\ee 
The gravitino mass breaks $SU(1,1)$ further to $U_a(1)$, since 
the K$\ddot{a}$hler function (\ref{37}) is not invariant under 
the dilatation subgroup.

\section{No--scale models with nontrivial superpotential and MPP}

The introduction of the superpotential complicates the analysis.
Suppose that the K$\ddot{a}$hler potential $K$ of the model is given
by Eq.~(\ref{37}) and the superpotential does not depend on the
singlet superfield $z$. Then one can define the vector $\alpha_i$
\be 
\alpha_i=e^{-K/3}\biggl[\ds\frac{1}{3}F_i(\varphi_{\alpha})-
\ds\frac{3+\sum_j
g_{\bar{j}}(\varphi_{\alpha},\bar{\varphi}_{\alpha})
F_{j}(\varphi_{\alpha})}{3|\partial_z f(z)|^2}f_i(z)
\biggr]\,, 
\label{312} 
\ee 
where $F(\varphi_{\alpha})=\ln W(\varphi_{\alpha})$ and the indices i and j on 
the functions $f(z)$, $g(\varphi_{\alpha},\bar{\varphi}_{\alpha})$ and 
$F(\varphi_{\alpha})$ denote the derivatives with respect to $z$ and $\varphi_{\alpha}$. 
The vector $\alpha_i$ satisfies the following property
$$
\sum_{j}G_{i\bar{j}}\alpha_j=G_i
$$
from which one deduces that 
\be
\sum_{i,\,k}G_{i}G^{i\bar{k}}G_{\bar{k}}=\sum_{k}\alpha_{k}G_{\bar{k}}\,.
\label{313} 
\ee 
As a result the scalar potential takes the form
\be 
V=\frac{1}{3}e^{2K/3}\sum_{\alpha}\biggl|\ds\frac{\partial
W(\varphi_{\alpha})}{\partial\varphi_{\alpha}}
\biggr|^2+\ds\frac{1}{2}\sum_{a}(D^{a})^2\,. 
\label{314}
\ee 
The potential (\ref{314}) leads to a supersymmetric particle
spectrum at low energies. It is positive definite and its minimum
is reached when $\biggl<\ds\frac{\partial
W(\varphi_{\alpha})}{\partial \varphi_{\alpha}}\biggr>=\,
<D^{a}>=0$, so that the cosmological constant goes to zero.

It is interesting to investigate what kind of symmetries protect
the cosmological constant when $W(z,\varphi_{\alpha})\ne const$.
As discussed above, it is natural to seek such symmetries within
the subgroups of $SU(1,1)$. The invariance of the K$\ddot{a}$hler
function under the imaginary translations of the hidden sector
superfields 
\be 
z_i\to z_i+i\beta_i\,;\qquad
\varphi_{\alpha}\to\varphi_{\alpha} 
\label{315} 
\ee 
implies that the K$\ddot{a}$hler potential depends only on 
$z_i+\bar{z}_i$, while the superpotential is given by 
\be
W(z_i,\varphi_{\alpha})=\exp\left\{\sum_{i=1}^{m} a_i
z_i\right\}\tilde{W}(\varphi_{\alpha})\,, 
\label{316} 
\ee 
where $a_i$ are real. Here we assume that the hidden sector involves $m$
singlet superfields. Since $G(\phi_M,\bar{\phi}_M)$ does not
change if
$$
\left\{
\ba{l}
K(\phi_M,\bar{\phi}_M)\to K(\phi_M,\bar{\phi}_M)-g(\phi_M)
-g^{*}(\bar{\phi}_M)\,,\\[2mm]
W(\phi_M)\to \ds e^{g(\phi_M)}W(\phi_M)
\ea
\right.\,.
$$
the most general K$\ddot{a}$hler function can be written as
\be
G(\phi_M,\bar{\phi}_M)=K(z_i+\bar{z}_i,\varphi_{\alpha},
\bar{\varphi}_{\alpha})+\ln|W(\varphi_{\alpha})|\,,
\label{318}
\ee
where $W(\varphi_{\alpha})=\tilde{W}(\varphi_{\alpha})$.

The dilatation invariance constrains the K$\ddot{a}$hler potential
and superpotential further. Suppose that hidden and observable
superfields transform differently 
\be 
z_i\to\alpha^k z_i\,,\qquad
\varphi_{\sigma}\to\alpha\varphi_{\sigma}\,. 
\label{319} 
\ee 
Then the superpotential $W(\varphi_{\alpha})$ may contain either bilinear or
trilinear terms involving the chiral superfields $\varphi_{\alpha}$ 
but not both. Because in phenomenologically acceptable
theories the masses of the observable fermions are generated by
trilinear terms, all others should be omitted. If there is only
one field $T$ in the hidden sector, then the K$\ddot{a}$hler function
is fixed uniquely by the gauge invariance and symmetry transformations 
(\ref{315}) and (\ref{319}): 
\be
\ba{c}
K(T+\bar{T},\varphi_{\sigma},\bar{\varphi}_{\sigma})=-\ds\frac{6}{k}\ln(T+\bar{T})+
\sum_{\sigma} C_{\sigma}\frac{|\varphi_{\sigma}|^2}{(T+\bar{T})^{2/k}}\\[2mm]
W(\varphi_{\alpha})=\ds\sum_{\sigma,\beta,\gamma}\ds\frac{1}{6}
Y_{\sigma\beta\gamma}\varphi_{\sigma}\varphi_{\beta}\varphi_{\gamma}\,, 
\ea 
\label{320}
\ee 
where $C_{\sigma}$ and $Y_{\sigma\beta\gamma}$ are constants.
The scalar potential of the hidden sector induced by the
K$\ddot{a}$hler function, with $K$ and $W$ given by Eq.~(\ref{320}),
is
$$
V(T+\bar{T})=\ds\frac{3}{(T+\bar{T})^{6/k}}\biggl[\ds\frac{2}{k}-1\biggr]
$$
and vanishes when $k=2$. In this case the subgroups of $SU(1,1)$
--- imaginary translations and dilatations ($T\to\alpha^2 T$,
$\varphi_{\sigma}\to \alpha\varphi_{\sigma}$) --- keep the value
of the cosmological constant equal to zero.

The invariance of the K$\ddot{a}$hler function with respect to
imaginary translations and dilatations prevents the breaking of
supersymmetry. In order to demonstrate this, let us consider the
SU(5) SUSY model with one field in the adjoint representation
$\Phi$ and with one singlet field $S$. The superpotential that
preserves gauge and global symmetries has the form 
\be
W(S,\Phi)=\ds\frac{\varkappa}{3}S^3+\lambda \mbox{Tr}\Phi^3+\sigma
S \mbox{Tr}\Phi^2\,. 
\label{321} 
\ee 
In the general case the
minimum of the scalar potential, which is induced by the
superpotential (\ref{321}), is attained when $<S>=<\Phi>=0$ and
does not lead to the breaking of local supersymmetry or of gauge
symmetry. But if $\varkappa=-40\sigma^3/(3\lambda^2)$ there is a
vacuum configuration \be <\Phi>=\ds\frac{\Phi_0}{\sqrt{15}}\left(
\ba{ccccc}
1 & 0 & 0 & 0 & 0 \\[0mm]
0 & 1 & 0 & 0 & 0 \\[0mm]
0 & 0 & 1 & 0 & 0 \\[0mm]
0 & 0 & 0 & -3/2 & 0 \\[0mm]
0 & 0 & 0 & 0 & -3/2
\ea
\right)\,,\qquad
\ba{l}
<S>=S_0\,, \\
\\
\Phi_0=\ds\frac{4\sqrt{15}\sigma}{3\lambda}S_0\,, 
\ea 
\label{322}
\ee 
which breaks SU(5) down to $SU(3)\times SU(2)\times U(1)$.
However, along the valley (\ref{322}), the superpotential and all
auxiliary fields $F_i$ vanish preserving supersymmetry and the
zero value of the vacuum energy density.

In order to get a vacuum where local supersymmetry is broken, one
should violate dilatation invariance, allowing the appearance of
the bilinear terms in the superpotential of SUGRA models.
Eliminating the singlet field from the considered SU(5) model and
introducing a mass term for the adjoint representation, we get 
\be
W(\Phi)=M_X\mbox{Tr}\Phi^2+\lambda \mbox{Tr}\Phi^3\,. 
\label{323}
\ee 
In the resulting model, there are a few degenerate vacua with
vanishing vacuum energy density. For example, in the scalar
potential there exists a minimum where $<\Phi>=0$ and another
vacuum, which has a configuration similar to Eq.~(\ref{322}) but
with $\Phi_0=\ds\frac{4\sqrt{15}}{3\lambda}M_X$. In the first
vacuum the SU(5) symmetry and local supersymmetry remain intact,
while in the second one the auxiliary field $F_{T}$ acquires a
vacuum expectation value and a non-zero gravitino mass is
generated: 
\be 
\ba{rclcl}
<|F_T|>&\simeq&\left<\ds\frac{|W(\Phi)|}{(T+\bar{T})^{1/2}}\right>&
=&m_{3/2}(T+\bar{T})\,,\\[3mm]
m_{3/2}&=&\left<\ds\frac{|W(\Phi)|}{(T+\bar{T})^{3/2}}\right>&
=&\ds\frac{40}{9}\frac{M_X^3}{\lambda^2(T+\bar{T})^{3/2}}\,. 
\ea
\label{324} 
\ee 
As a result, local supersymmetry and gauge symmetry are broken in the second vacuum. 
However it does not break global supersymmetry in the observable sector 
at low energies (see Eq.(\ref{314})). When $M_X$ goes to zero
the dilatation invariance, SU(5) symmetry and local supersymmetry
are restored.

A simple model with the superpotential (\ref{323}) can serve as a
basis for the Multiple Point Principle (MPP) assumption in  SUGRA
models, which was formulated recently in \cite{37}. When applied
to supergravity, MPP implies that the scalar potential contains at
least two degenerate minima. In one of them local supersymmetry is
broken in the hidden sector, inducing a set of soft SUSY breaking
terms for the observable fields. In the other vacuum the low
energy limit of the considered theory is described by a pure
supersymmetric model in flat Minkowski space. Since the vacuum
energy density of supersymmetric states in flat Minkowski space is
just zero, the cosmological constant problem is thereby solved to
first approximation by the MPP assumption. An important point is
that the vacua with broken and unbroken local supersymmetry are
degenerate and have zero energy density in the model considered
above. However, in the vacuum where local supersymmetry is broken,
all soft SUSY breaking terms vanish making this model irrelevant
for phenomenological studies.

\section{No--scale models and the superstring}

The K$\ddot{a}$hler function and the structure of the hidden sector
should be fixed by an underlying renormalizable or even finite
theory. Nowadays the best candidate for the ultimate theory is
$E_8\times E_8$ (ten dimensional) heterotic superstring theory
\cite{48}. The minimal possible SUSY--breaking sector in string
models involves dilaton ($S$) and moduli ($T_m$) superfields. The
number of moduli varies from one string model to another. But
dilaton and moduli fields are always present in four--dimensional
heterotic superstrings, because $S$ is related with the
gravitational sector while vacuum expectation values of $T_m$
determine the size and shape of the compactified space. Amongst
the moduli $T_m$ we concentrate here on the overall modulus $T$.
In this case Calabi--Yau and orbifold compactifications lead to
rather similar results for the K$\ddot{a}$hler potential,
superpotential and gauge kinetic functions at the tree level: 
\be
\ba{c}
K=-\ln(S+\bar{S})-3\ln(T+\bar{T})+\sum_{\alpha}(T+\bar{T})^{n_{\alpha}}
\varphi_{\alpha}\bar{\varphi}_{\alpha}\,,\\[2mm]
W=W^{(ind)}(S,\,T,\,\varphi_\alpha)+\ds\sum_{\sigma,\beta,\gamma}
\ds\frac{1}{6} Y_{\sigma\beta\gamma}\varphi_{\sigma}
\varphi_{\beta}\varphi_{\gamma}\,,\qquad f_a=k_a S\,, 
\ea
\label{41} 
\ee 
where $k_a$ is the Kac--Moody level of the gauge
factor ($k_3=k_2=\ds\frac{3}{5}k_1=1$). In the case of orbifold
compactifications, the $n_{\alpha}$ are negative integers
sometimes called modular weights of the matter fields. Orbifold
models have a symmetry (``target--space duality'') which is either
the modular group $SL(2,\bf{Z})$ or a subgroup of it. Under
$SL(2,\bf{Z})$, the fields transform like 
\be 
\ba{c}
T\to\ds\frac{aT-ib}{icT+d}\,,\qquad ad-bc=1\,
\quad a,\,b,\,c,\,d\,\in\, \bf{Z};\\[2mm]
S\to S\,;\qquad\varphi_{\alpha}\to(icT+
d)^{n_{\alpha}}\varphi_{\alpha}\,. 
\ea 
\label{42} 
\ee 
In the large $T$ limit of the Calabi--Yau compactifications, 
$n_{\alpha}=-1$ and the Lagrangian of the effective SUGRA models 
is also invariant with respect to the field transformations (\ref{42}) 
if $n_{\alpha}=-1$. So one can see that the form of the
K$\ddot{a}$hler function is very close to the no--scale structure
discussed in the previous sections.

In the classical limit $W^{(ind)}(S,\,T,\,\varphi_\alpha)$ is
absent. The superpotential of the hidden sector and supersymmetric
mass terms of the observable superfields may be induced by
non-perturbative corrections, which violate the invariance under
$SL(2,\bf{Z})$ symmetry. In the gaugino condensation scenario for
SUSY breaking, the superpotential of the hidden sector takes the
form: \be W(S,\,T)\sim \exp\left\{-3S/2b_Q\right\}\,, \label{43}
\ee where $b_{Q}$ is the beta--function of the hidden sector gauge
group. For an $SU(N)$ model without matter superfields
$b_{Q}=3N/(16\pi^2)$. Assuming that the superpotential does not
depend on $T$, we get 
\be
V(S,\,T)=\ds\frac{1}{(S+\bar{S})(T+\bar{T})^3}\biggl|\frac{\partial
W(S)}{\partial S}-\frac{W(S)}{S+\bar{S}}\biggr|^2\,. 
\label{44}
\ee 
The scalar potential (\ref{44}) of the hidden sector is
positive definite. All its vacua are degenerate and have zero
energy density. Among them there can be a minimum where the vacuum
expectation value of the hidden sector superpotential vanishes. It
is easy to check that, in this vacuum, local supersymmetry remains
intact. In other vacua where $<W(S)>\ne 0$ local supersymmetry is
broken, since $F_{T}\ne 0$. Thus the MPP conditions can be
realized in superstring inspired models as well.

But at low energies the SUGRA Lagrangian, corresponding to the
K$\ddot{a}$hler function given by Eq.~(\ref{41}) with $n_{\alpha}=-1$
and a superpotential that does not depend on the overall modulus
$T$, exhibits structure inherent in global supersymmetry. In order
to destroy the degeneracy between bosons and fermions, the
$SL(2,\bf{Z})$ symmetry should be broken further. Non-zero gaugino
masses $M_a$ are generated when the gauge kinetic function gets a
dependence on $T$, i.e. $f_a=k_a (S-\sigma T)$. The soft scalar
masses $m_{\alpha}^2$ and trilinear couplings
$A_{\alpha\beta\gamma}$ arise for the minimal choice of the
K$\ddot{a}$hler metric of the observable superfields, when the
K$\ddot{a}$hler potential is given by 
\be
K=-\ln(S+\bar{S})-3\ln(T+\bar{T})+
\sum_{\alpha}\varphi_{\alpha}\bar{\varphi}_{\alpha}\,. 
\label{45}
\ee 
In this case we have 
\be 
A_{\alpha\beta\gamma}=3m_{3/2}\,,\qquad
m_{\alpha}^2=m_{3/2}^2, 
\label{46} 
\ee 
It is worth emphasizing that
the energy densities of vacua still vanish in models with the
modified gauge kinetic function and K$\ddot{a}$hler potential
(\ref{45}). It clears the way to the construction of realistic
SUGRA models based on the MPP assumption.

\vspace{-2mm}
\section*{Acknowledgements}
\vspace{-2mm}
The authors are grateful to E.I.~Guendelman and
N.S.~Mankoc-Borstnik for stimulating questions and comments, and
S.F.~King, O.V.Kancheli, S.~Moretti, M.~Sher and
M.I.~Vysotsky for fruitful discussions. The work of RN was
supported by the Russian Foundation for Basic Research (projects
00-15-96562 and 02-02-17379) and by a Grant of the President of
Russia for young scientists (MK--3702.2004.2).


\vspace{-3mm}
\begin{thebibliography}{99}
\vspace{-2mm}
\bibitem{0} A.G.Riess {\itshape et al.}, Astron.J. {\bf 116}, 1009 (1998);
S.Perlmutter {\itshape et al.}, Astrophys.J. {\bf 517}, 565 (1999);
C.Bennett {\itshape et al.}, astro-ph/0302207;
D.Spergel {\itshape et al.}, astro-ph/0302209.
\vspace{-2mm}
\bibitem{01}
J.Polonyi, Budapest preprint KFKI--1977 93 (1977).
\vspace{-2mm}
\bibitem{34}
A.B.Lahanas, D.V.Nanopoulos, Phys.Rep. 145 (1987) 1.
\vspace{-2mm}
\bibitem{02}
D.L.Bennett, H.B.Nielsen, Int.J.Mod.Phys. A {\bf 9}, 5155 (1994); {\it ibid} {\bf 14}, 3313 (1999);
D.L.Bennett, C.D.Froggatt, H.B.Nielsen, in {\itshape Proceedings of the 27th International Conference on
High energy Physics, Glasgow, Scotland, 1994}, p.557;
{\itshape Perspectives in Particle Physics '94, World Scientific, 1995}, p. 255, ed. D. Klabu\u{c}ar,
I. Picek and D. Tadi\'{c} [arXiv:hep-ph/9504294].
\vspace{-2mm}
\bibitem{21}
H.P.Nilles, Phys.Rep. {\bf 110}, 1 (1984).
\vspace{-2mm}
\bibitem{31}
E.Gremmer, S.Ferrara, C.Kounnas, D.V.Nanopoulos, Phys.Lett. B 133 (1983) 61.
\vspace{-2mm}
\bibitem{32}
J.Ellis, M.K.Gaillard, M.G$\ddot{u}$naydin, B.Zumino, Nucl.Phys. B 224 (1983) 427.
\vspace{-2mm}
\bibitem{33}
J.Ellis, C.Kounnas, D.V.Nanopoulos, Nucl.Phys. B 241 (1984) 406.
\vspace{-2mm}
\bibitem{35}
J.Ellis, C.Kounnas, D.V.Nanopoulos, Nucl.Phys. B 247 (1984) 373.
\vspace{-2mm}
\bibitem{36}
E.Gremmer, B.Julia, Phys.Lett.B 80 (1978) 48; Nucl.Phys.B 159 (1979) 141.
\vspace{-2mm}
\bibitem{37}
C.Froggatt, L.Laperashvili, R.Nevzorov, H.B.Nielsen, Phys.Atom.Nucl.67 (2004) 582.
\vspace{-2mm}
\bibitem{48} 
M.B.Green, J.H.Schwarz, E.Witten, {\itshape Superstring Theory, Cambridge Univ. Press, Cambridge, 1987}.

\end{thebibliography}
\end{document}